\documentclass[aps,prd,preprint,tightenlines,superscriptaddress,showpacs,byrevtex]{revtex4}
\usepackage{graphicx}
\usepackage{dcolumn}
\usepackage{bm}

\newcommand*{\sinbb}{{\sin2\phi_1}}
\newcommand*{\dM}{\ensuremath{{\Delta m_d}}}
\newcommand*{\dmd}{\dM}

\newcommand*{\bz}{\ensuremath{{B^0}}}
\newcommand*{\bzb}{\ensuremath{{\overline{B}{}^0}}}
\newcommand*{\bp}{\ensuremath{{B^+}}}

\newcommand*{\piz}{\ensuremath{{\pi^0}}}
\newcommand*{\pip}{\ensuremath{{\pi^+}}}
\newcommand*{\pim}{\ensuremath{{\pi^-}}}
\newcommand*{\kp}{\ensuremath{{K^+}}}
\newcommand*{\km}{\ensuremath{{K^-}}}
\newcommand*{\ks}{\ensuremath{{K_S^0}}}
\newcommand*{\kl}{\ensuremath{{K_L^0}}}
\newcommand*{\kstar}{\ensuremath{{K^{*}}}}
\newcommand*{\kstarz}{\ensuremath{{K^{*0}}}}
\newcommand*{\jpsi}{\ensuremath{{J/\psi}}}
\newcommand*{\dm}{\ensuremath{{D^-}}}
\newcommand*{\dsm}{\ensuremath{{D^{*-}}}}
\newcommand*{\rhop}{\ensuremath{{\rho^+}}}
\newcommand*{\dzb}{\ensuremath{{\overline{D}{}^0}}}

\newcommand*{\Dt}{\ensuremath{{\Delta t}}}
\newcommand*{\Dz}{\ensuremath{{\Delta z}}}
\def\dt{\Dt}

\newcommand*{\fcp}{\ensuremath{{f_{CP}}}}
\newcommand*{\ftag}{\ensuremath{{f_\mathrm{tag}}}}
\newcommand*{\fflv}{\ensuremath{{f_\mathrm{flv}}}}

\newcommand*{\tcp}{\ensuremath{{t_{CP}}}}
\newcommand*{\ttag}{\ensuremath{{t_\mathrm{tag}}}}

\newcommand*{\zcp}{\ensuremath{{z_{CP}}}}
\newcommand*{\ztag}{\ensuremath{{z_\mathrm{tag}}}}

\newcommand*{\dE}{\ensuremath{{\Delta E}}}
\newcommand*{\mb}{\ensuremath{{M_\mathrm{bc}}}}
\newcommand*{\Ebeam}{\ensuremath{{E_\mathrm{beam}^\mathrm{cms}}}}
\newcommand*{\EB}{\ensuremath{{E_B^\mathrm{cms}}}}
\newcommand*{\pB}{\ensuremath{{p_B^\mathrm{cms}}}}

\newcommand*{\Nev}{\ensuremath{{N_\mathrm{ev}}}}

\newcommand*{\eeff}{\ensuremath{\epsilon_\mathrm{eff}}}

\newcommand*{\taubz}{\ensuremath{{\tau_\bz}}}
\newcommand*{\taubp}{\ensuremath{{\tau_\bp}}}
\newcommand*{\taubppr}{\ensuremath{{\tau'_\bp}}}
\def\taub{\taubz}
\newcommand*{\taubratio}{\ensuremath{{\taubp/\taubz}}}

\newcommand*{\Psig}{\ensuremath{{\mathcal{P}_\mathrm{sig}}}}
\newcommand*{\Pbkg}{\ensuremath{{\mathcal{P}_\mathrm{bkg}}}}
\newcommand*{\Pol}{\ensuremath{{P_\mathrm{ol}}}}
\newcommand*{\Rsig}{\ensuremath{{R_\mathrm{sig}}}}
\newcommand*{\Rbkg}{\ensuremath{{R_\mathrm{bkg}}}}
\newcommand*{\fol}{\ensuremath{{f_\mathrm{ol}}}}
\newcommand*{\fsig}{\ensuremath{{f_\mathrm{sig}}}}

\newcommand*{\dwl}{\ensuremath{{\Delta w_l}}}
\newcommand*{\fq}{\ensuremath{q}}

\newcommand{\cala}{{\cal A}}
\newcommand{\cals}{{\cal S}}

\newcommand*{\sinbbcenter}{0.728}
\newcommand*{\sinbbstat}{0.056}
\newcommand*{\sinbbsys}{0.023}
\newcommand*{\sinbbresult}{\sinbbcenter\pm\sinbbstat\mathrm{(stat)}\pm\sinbbsys\mathrm{(syst)}}
\newcommand*{\efftot}{0.287 \pm 0.005}
\newcommand*{\lambdacenter}{1.007}
\newcommand*{\lambdastat}{0.041}
\newcommand*{\lambdasys}{0.033}
\newcommand*{\lambdaresult}{\lambdacenter\pm\lambdastat\mathrm{(stat)}\pm\lambdasys\mathrm{(syst)}}

\newcommand*{\nevqp}{2717}
\newcommand*{\nevqm}{2700}

\newcommand{\mdiff}{M_{\rm diff}}
\newcommand{\dslnu}{D^{*-}\ell^+\nu}
\newcommand{\bzdslnu}{\bz \to \dslnu}

\newcommand{\kppm}{K^+\pi^-}
\newcommand{\kppmpz}{\kppm\pi^0}
\newcommand{\kppmpppm}{\kppm\pi^+\pi^-}
\newcommand{\dzbkppm}{\dzb \to \kppm}
\newcommand{\dzbkppmpz}{\dzb \to \kppmpz}
\newcommand{\dzbkppmpppm}{\dzb \to \kppmpppm}
\newcommand{\thetabdl}{\theta_{B,D^*\ell}}
\newcommand{\cosbdl}{\cos\thetabdl}
\newcommand{\mnu}{M_\nu}

\newcommand{\taubzcenter}{1.534}
\newcommand{\taubzstat}{0.008}
\newcommand{\taubzsys}{0.010}
\newcommand{\taubzresult}{[\taubzcenter\pm\taubzstat\mathrm{(stat)}\pm\taubzsys\mathrm{(syst)]~ps}}
\newcommand{\taubpcenter}{1.635}
\newcommand{\taubpstat}{0.011}
\newcommand{\taubpsys}{0.011}
\newcommand{\taubpresult}{[\taubpcenter\pm\taubpstat\mathrm{(stat)}\pm\taubpsys\mathrm{(syst)]~ps}}

\newcommand{\taubratiocenter}{1.066}
\newcommand{\taubratiostat}{0.008}
\newcommand{\taubratiosys}{0.008}
\newcommand{\taubratioresult}{\taubratiocenter\pm\taubratiostat\mathrm{(stat)}\pm\taubratiosys\mathrm{(syst)}}

\newcommand{\dmdcenter}{0.511}
\newcommand{\dmdstat}{0.005}
\newcommand{\dmdsys}{0.006}
\newcommand{\dmdresult}{[\dmdcenter\pm\dmdstat\mathrm{(stat)}\pm\dmdsys
            \mathrm{(syst)]~ps}^{-1}}

\def\ufours{\Upsilon(4S)}

\begin{document}

\vspace*{-3\baselineskip}
\resizebox{!}{3cm}{\includegraphics{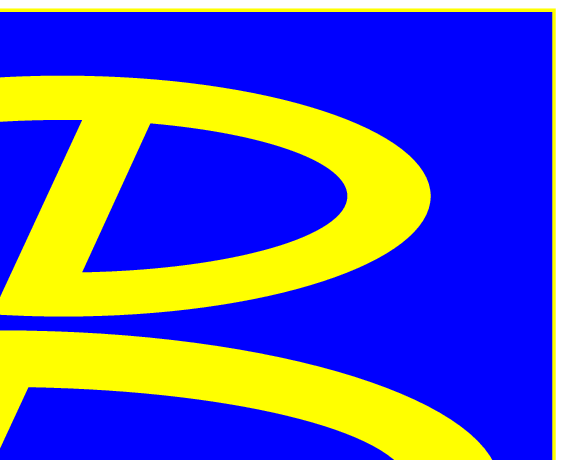}}

\vspace*{-3cm}
\begin{flushright}
BELLE-CONF-0436\\
Belle Preprint 2004-31\\
KEK   Preprint 2004-63
\end{flushright}
\vspace*{1cm}

\title{Improved Measurement of 
{\boldmath $CP$}-Violation Parameters {\boldmath $\sin2\phi_1$} and
{\boldmath $|\lambda|$},
{\boldmath $B$} Meson Lifetimes, and
{\boldmath $\bz$}-{\boldmath $\bzb$} Mixing Parameter {\boldmath $\dM$}
}

\date{\today}
\affiliation{Budker Institute of Nuclear Physics, Novosibirsk}
\affiliation{Chiba University, Chiba}
\affiliation{Chonnam National University, Kwangju}
\affiliation{University of Cincinnati, Cincinnati, Ohio 45221}
\affiliation{University of Frankfurt, Frankfurt}
\affiliation{Gyeongsang National University, Chinju}
\affiliation{University of Hawaii, Honolulu, Hawaii 96822}
\affiliation{High Energy Accelerator Research Organization (KEK), Tsukuba}
\affiliation{Hiroshima Institute of Technology, Hiroshima}
\affiliation{Institute of High Energy Physics, Chinese Academy of Sciences, Beijing}
\affiliation{Institute of High Energy Physics, Vienna}
\affiliation{Institute for Theoretical and Experimental Physics, Moscow}
\affiliation{J. Stefan Institute, Ljubljana}
\affiliation{Kanagawa University, Yokohama}
\affiliation{Korea University, Seoul}
\affiliation{Kyungpook National University, Taegu}
\affiliation{Swiss Federal Institute of Technology of Lausanne, EPFL, Lausanne}
\affiliation{University of Ljubljana, Ljubljana}
\affiliation{University of Maribor, Maribor}
\affiliation{University of Melbourne, Victoria}
\affiliation{Nagoya University, Nagoya}
\affiliation{Nara Women's University, Nara}
\affiliation{National Central University, Chung-li}
\affiliation{National United University, Miao Li}
\affiliation{Department of Physics, National Taiwan University, Taipei}
\affiliation{H. Niewodniczanski Institute of Nuclear Physics, Krakow}
\affiliation{Nihon Dental College, Niigata}
\affiliation{Niigata University, Niigata}
\affiliation{Osaka City University, Osaka}
\affiliation{Osaka University, Osaka}
\affiliation{Panjab University, Chandigarh}
\affiliation{Peking University, Beijing}
\affiliation{Princeton University, Princeton, New Jersey 08545}
\affiliation{Saga University, Saga}
\affiliation{University of Science and Technology of China, Hefei}
\affiliation{Seoul National University, Seoul}
\affiliation{Sungkyunkwan University, Suwon}
\affiliation{University of Sydney, Sydney NSW}
\affiliation{Tata Institute of Fundamental Research, Bombay}
\affiliation{Toho University, Funabashi}
\affiliation{Tohoku Gakuin University, Tagajo}
\affiliation{Tohoku University, Sendai}
\affiliation{Department of Physics, University of Tokyo, Tokyo}
\affiliation{Tokyo Institute of Technology, Tokyo}
\affiliation{Tokyo Metropolitan University, Tokyo}
\affiliation{Tokyo University of Agriculture and Technology, Tokyo}
\affiliation{University of Tsukuba, Tsukuba}
\affiliation{Virginia Polytechnic Institute and State University, Blacksburg, Virginia 24061}
\affiliation{Yonsei University, Seoul}
  \author{K.~Abe}\affiliation{High Energy Accelerator Research Organization (KEK), Tsukuba} 
  \author{K.~Abe}\affiliation{Tohoku Gakuin University, Tagajo} 
  \author{I.~Adachi}\affiliation{High Energy Accelerator Research Organization (KEK), Tsukuba} 
  \author{H.~Aihara}\affiliation{Department of Physics, University of Tokyo, Tokyo} 
  \author{M.~Akatsu}\affiliation{Nagoya University, Nagoya} 
  \author{Y.~Asano}\affiliation{University of Tsukuba, Tsukuba} 
  \author{T.~Aushev}\affiliation{Institute for Theoretical and Experimental Physics, Moscow} 
  \author{T.~Aziz}\affiliation{Tata Institute of Fundamental Research, Bombay} 
  \author{S.~Bahinipati}\affiliation{University of Cincinnati, Cincinnati, Ohio 45221} 
  \author{A.~M.~Bakich}\affiliation{University of Sydney, Sydney NSW} 
% ...
\collaboration{The Belle Collaboration}


\begin{abstract}
We present a precise measurement of the standard model $CP$-violation
parameter $\sinbb$, the direct $CP$ violation parameter $|\lambda|$,
the lifetimes of charged and neutral $B$ mesons and their ratio,
and the $\bz$-$\bzb$ mixing parameter $\dM$ based on a sample
of $152 \times 10^6$ $B\overline{B}$ pairs
collected at the $\Upsilon(4S)$ resonance
with the Belle detector at the KEKB asymmetric-energy $e^+e^-$ collider.
One of two $B$ mesons is fully reconstructed in a $CP$-eigenstate
or a flavor-eigenstate decay channel.
The flavor of the accompanying $B$ meson is
identified from its decay products.
From the distributions of the time interval
between the two $B$ meson decay points, we obtain
 $\sinbb     = \sinbbresult$,
 $|\lambda|  = \lambdaresult$,
 $\taubz     = \taubzresult$,
 $\taubp     = \taubpresult$,
 $\taubratio = \taubratioresult$ and
 $\dmd       = \dmdresult$.
The results for $\sinbb$ and $|\lambda|$
are consistent with the standard model expectations.
The significance of
the observed deviation from unity in the lifetime ratio 
exceeds five standard deviations.
\end{abstract}


\maketitle

\section{Introduction}
\label{sec:introduction}
In the standard model (SM), $CP$ violation arises from an
irreducible phase in the weak interaction quark-mixing matrix
[Cabibbo-Kobayashi-Maskawa (CKM) matrix]~\cite{bib:ckm}.
In particular, the SM predicts a $CP$-violating asymmetry
in the time-dependent rates for $\bz$ and $\bzb$ decays
to a common $CP$ eigenstate $\fcp$,
where the transition is dominated by the $b \to c\overline{c}s$ process,
with negligible corrections from strong interactions~\cite{bib:sanda}:
\begin{equation}
  A(t) \equiv \frac{\Gamma[\bzb(t) \to \fcp] - \Gamma[\bz(t) \to \fcp]}
  {\Gamma[\bzb(t) \to \fcp] + \Gamma[\bz(t) \to \fcp]}
  = -\xi_f \sinbb \sin(\dM t),
\end{equation}
where $\Gamma[\bz(t),\bzb(t) \to \fcp]$ is the rate for $\bz$ or $\bzb$
decay to $\fcp$ at a proper time $t$ after production,
$\xi_f$ is the $CP$ eigenvalue of $\fcp$,
$\dM$ is the mass difference between the two $\bz$ mass eigenstates,
and $\phi_1$ is one of the three interior angles of the CKM unitarity triangle,
defined as $\phi_1 \equiv \pi - \arg(V_{tb}^*V_{td}/V_{cb}^*V_{cd})$.
Non-zero values for $\sinbb$ have been reported
by the Belle and BaBar
collaborations~\cite{bib:cpv,bib:Belle_sin2phi1_78fb-1,bib:babar}.
Belle's latest published measurement of $\sinbb$ is based on
a 78~fb$^{-1}$ data sample (data set I) containing $85 \times 10^{6}$
$B\overline{B}$ pairs produced at the $\Upsilon(4S)$ resonance.
In this paper, we report an improved measurement 
incorporating an additional 62 fb$^{-1}$ (data set II) for a total of
140 fb$^{-1}$ ($152 \times 10^6$ $B\overline{B}$ pairs).
A precise knowledge of $\sin 2\phi_1$ is essential for testing the
Kobayashi-Maskawa model of $CP$ violation.

The $\sinbb$ measurement requires a determination of
a proper-time resolution function and of the wrong-tag
fractions using a large sample of exclusively reconstructed
flavor-eigenstate decays.
We perform a precise measurement of the mixing parameter $\dmd$ and
of the neutral (charged) $B$ meson lifetime $\taubz$ ($\taubp$) as a 
byproduct of this procedure.
Our previous results are based on
a 29.1 fb$^{-1}$ data sample~\cite{Abe:2002id,Tomura:2002qs,Hara:2002mq}; thus
our new measurements with a 140 fb$^{-1}$ data sample
provide significant improvements.

Changes exist in the analysis
with respect to our earlier results.
We apply a new proper-time resolution function
that reduces systematic uncertainties.
We introduce $b$-flavor-dependent wrong-tag fractions
to accommodate possible differences between $\bz$ and $\bzb$ decays.
We also adopt a multi-parameter fit to the flavor-eigenstate samples
to obtain $\dmd$, $\taubz$, $\taubp$,
the resolution parameters and wrong-tag fractions simultaneously.
There are other improvements in the estimation of background
components that are made possible by the increased statistics.

The data were collected with the Belle detector~\cite{bib:belle}
at the KEKB  asymmetric-energy $e^+e^-$ collider~\cite{bib:KEKB},
which collides 8.0~GeV $e^-$ on 3.5~GeV $e^+$
at a small ($\pm 11$~mrad) crossing angle.
We use events where one of the $B$ mesons decays to $\fcp$ at time $\tcp$,
and the other decays to a self-tagging state $\ftag$,
which distinguishes $\bz$ from $\bzb$, at time $\ttag$.
The $CP$ violation manifests itself as an asymmetry $A(\Dt)$,
where $\Dt$ is the proper time interval
between the two decays: $\Dt \equiv \tcp - \ttag$.
We also use events in which $\fcp$ is replaced by a flavor eigenstate $\fflv$;
the decay chain in this case is $\ufours\to\bz\bzb\to\fflv\ftag$.
The time evolution is described as
${e^{-|\Dt|/\taubz}}/(4\taubz)\{1 \pm \cos (\dmd \Delta t)\}$, where
the plus (minus) sign is taken when the
flavor of one $B$ meson is opposite to (the same as) the other.

At KEKB, the $\Upsilon(4S)$ resonance is produced
with a boost of $\beta\gamma = 0.425$ nearly along the $z$ axis
defined as anti-parallel to the positron beam direction,
and $\Dt$ can be determined as $\Dt \simeq \Dz/(\beta\gamma c)$,
where $\Dz$ is the $z$ distance between the $\fcp$ and $\ftag$
decay vertices, $\Dz \equiv \zcp - \ztag$.
The average value of $\Dz$ is approximately 200~$\mu$m.

The Belle detector~\cite{bib:belle} is a large-solid-angle spectrometer
that includes a silicon vertex detector (SVD),
a central drift chamber (CDC),
an array of aerogel threshold \v{C}erenkov counters (ACC),
time-of-flight (TOF) scintillation counters,
and an electromagnetic calorimeter (ECL) comprised of CsI(Tl) crystals,
all located inside a superconducting solenoid coil
that provides a 1.5~T magnetic field.
An iron flux-return located outside of the coil is instrumented
to detect $\kl$ mesons and to identify muons (KLM).

\section{Event selection and reconstruction}
\label{sec:selection}

\subsection{Reconstruction of 
            {\boldmath $\bz \to {\rm charmonium}~K^{(*)0}$} decays}
\label{sec:selection:ccs}
We reconstruct $\bz$ decays to the following $CP$ 
eigenstates~\cite{footnote:cc}:
$\jpsi\ks$, $\psi(2S)\ks$, $\chi_{c1}\ks$, $\eta_c\ks$ for $\xi_f = -1$
and $\jpsi\kl$ for $\xi_f = +1$.
We also use $\bz \to \jpsi\kstarz$ decays with the 
subsequent decay $\kstarz \to \ks\piz$.
Here the final state is a mixture of even and odd $CP$,
depending on the relative orbital angular momentum of the $\jpsi$ and $\kstarz$.
We find that the final state is primarily $\xi_f = +1$;
the $\xi_f = -1$ fraction is
$0.19 \pm 0.02 \mathrm{(stat)} \pm 0.03 \mathrm{(syst)}$~\cite{bib:itoh}.

The reconstruction and selection criteria for all $\fcp$ channels
used in the measurement are described in detail elsewhere~\cite{bib:cpv}.
$\jpsi$ and $\psi(2S)$ mesons are reconstructed
via their decays to $\ell^+\ell^-$ ($\ell = \mu,e$).
The $\psi(2S)$ is also reconstructed via $\jpsi\pip\pim$,
and the $\chi_{c1}$ via $\jpsi\gamma$.
The $\eta_c$ is detected in the $\ks\km\pip$, $\kp\km\piz$,
and $p\overline{p}$ modes.
For the $\jpsi\ks$ mode, we use $\ks \to \pip\pim$ and $\piz\piz$ decays;
for other modes we only use $\ks \to \pip\pim$.
For reconstructed $B \to \fcp$ candidates other than $\jpsi\kl$,
we identify $B$ decays using the energy difference $\dE \equiv \EB - \Ebeam$
and the beam-energy constrained mass $\mb \equiv \sqrt{(\Ebeam)^2-(\pB)^2}$,
where $\Ebeam$ is the beam energy in the center-of-mass system (cms)
of the $\Upsilon(4S)$ resonance, and $\EB$ and $\pB$ are
the cms energy and momentum of the reconstructed $B$ candidate, respectively.

Candidate $\bz \to \jpsi\kl$ decays are selected by requiring
ECL and/or KLM hit patterns that are consistent with the presence
of a shower induced by a $\kl$ meson.
The centroid of the shower is required to be within a $45^\circ$ cone
centered on the $\kl$ direction inferred from
two-body decay kinematics and the measured four-momentum of the $\jpsi$.

\subsection{Reconstruction of flavor-eigenstate samples}
\label{sec:selection:control}

\subsubsection{$\bz\to\dslnu$}
\label{sec:selection:control:dslnu}
We use the decay chain
$B^{0} \to D^{*-} \ell^{+} \nu$, $D^{*-} \to \dzb \pi^-$, where
$\dzbkppm$, $\kppmpz$ or $\kppmpppm$.
We require
associated SVD hits and radial
impact parameters $dr <0.2$~cm for all tracks.
Track momenta in the laboratory frame
for $\dzbkppmpppm$ decays are required to be larger than 0.2 GeV/$c$,
while no additional requirements are applied for the other modes.
Charged kaons are identified by combining
information from the TOF, ACC and $dE/dx$
measurements in the CDC.
Photon candidates are defined as isolated ECL clusters of more than
20~MeV that are not matched to any charged track.
$\piz$ candidates are reconstructed from pairs of photon candidates
with invariant masses between 124 and 146 MeV$/c^2$.
A mass-constrained fit is performed to improve the $\piz$ momentum resolution.
A minimum $\piz$ momentum of 0.2 GeV$/c$ is required.
For $\dzbkppm$ and $\kppmpppm$ candidates, we
use daughter combinations that have an invariant mass
within $0.013~{\rm GeV}/c^2$ of $m_{D^0}$;
for $\dzbkppmpz$ we expand 
the mass window to
$-0.037~{\rm GeV}/c^2$ and $+0.023~{\rm GeV}/c^2$.
For $D^{*-} \to \dzb \pi^-$ decays,
we combine $\dzb$ candidates with a low-momentum $\pi^-$
(slow pion) that is reconstructed using a vertex constraint
and require the mass difference between the $D^{*-}$ and
$\dzb$ candidates, $\mdiff$, to be within 1 MeV/$c^2$ of the nominal value.
We reject $D^{*-}$ candidates with cms momentum greater
than 2.6~GeV/$c$, which is beyond the kinematic limit for
$B$ meson decays.

For the associated lepton, we use electrons or muons 
with a charge opposite to that of the $D^{*-}$ candidate.
Electron identification is based on a combination of
CDC $dE/dx$ information, the ACC response, and the energy deposition of
the associated ECL shower. Muons are identified by comparing information
from the KLM to extrapolated charged particle trajectories.
We require $1.4~\text{GeV/}c <p^{\rm cms}_{\ell}<2.4$ GeV/$c$,
where $p^{\rm cms}_{\ell}$ is the cms momentum of the lepton.
The cms angle of the lepton
with respect to the direction of the $D^{*-}$ candidate
is also required to be greater than 90 degrees.
For $\bzdslnu$ decays, the energies and momenta
of the $B$ meson and the $D^*\ell$ system in the cms satisfy
$\mnu^2 = (E^{\rm cms}_B-E^{\rm cms}_{D^*\ell})^2-
|\vec{p}^{\rm ~cms}_B|^2-|\vec{p}^{\rm ~cms}_{D^*\ell}|^2+
2|\vec{p}^{\rm ~cms}_B|\,|\vec{p}^{\rm ~cms}_{D^*\ell}|\cosbdl$,
where $\mnu$ is the neutrino mass and $\thetabdl$ is the
angle between
$\vec{p}^{\rm ~cms}_B$ and $\vec{p}^{\rm ~cms}_{D^*\ell}$.
We calculate $\cosbdl$ setting $\mnu=0$.
The signal region is defined as $|\cosbdl|<1.1$.
We also require the candidate $\dslnu$ decays to be
outside the signal region when we artificially reverse the
lepton momentum vector; this ensures that
entries in the signal region after the reversal,
which are used for the background estimation,
do not contain $\dslnu$ events.

\subsubsection{Hadronic modes}
$\bz$ and $\bp$ mesons are fully reconstructed in the following decay modes:
$\bz \to \dm\pip$, $\dsm\pip$, $\dsm\rhop$,
$\jpsi\kstarz$, $\bp \to \dzb\pip$, and $\jpsi\kp$.
We also use $\bz\to\jpsi\ks$ decays, which are described in
the previous section, with no flavor assignment.
$\dsm$ and $\dzb$ candidates are reconstructed in the same decay modes 
that are used for the  $\dslnu$ mode.
Charged $D$ candidates are reconstructed in the $\dm\to\kp\pim\pim$ channel.
For $D$ and $\dsm$ candidates, we apply mode-dependent requirements
on the reconstructed $D$ mass (ranging from $\pm 15$~MeV$/c^2$ to $\pm
50$~MeV$/c^2$) 
and the $\mdiff$ (ranging from $\pm 3$~MeV$/c^2$ to $\pm 12$~MeV$/c^2$),
in a similar way as for
the $\dslnu$ mode.
%
Candidate $\kstarz\to \kp\pim$ decays are
required to have an invariant mass within 75 MeV$/c^2$
of the nominal $\kstarz$ mass.
$\rhop$ candidates are selected as $\pip\piz$ pairs having invariant masses
within 150 MeV$/c^2$ of the nominal $\rhop$ mass.
%

To reduce background from $e^+e^- \rightarrow q\bar{q}$
($q = u,d,s$~or~$c$) continuum events,
a selection based on the ratio of the second to 
zeroth Fox-Wolfram moments~\cite{Fox:1978vu}
and the angle between the thrust axes of the reconstructed and
associated $B$ mesons is applied mode by mode.

\subsection{Flavor tagging}
\label{sec:selection:fbtg}

For neutral $B$ to $\fcp$ and $\fflv$ decays, 
charged leptons, pions, kaons, and $\Lambda$ baryons
that are not associated with the reconstructed decay
are used to identify the $b$-flavor of the accompanying $B$ meson.
The tagging algorithm is described in detail 
elsewhere~\cite{Kakuno:2004cf}.
We use two parameters, $\fq$ and $r$, to represent the tagging information.
The first, $q$, has the discrete value $+1$~($-1$)
when the tag-side $B$ meson is likely to be a $\bz$~($\bzb$).
The parameter $r$ corresponds to an event-by-event 
flavor-tagging dilution that ranges
from $r=0$ for no flavor discrimination
to $r=1$ for an unambiguous flavor assignment.
It is determined from
a large number of events generated by Monte Carlo (MC) simulation,
and is used only to sort data into six intervals of $r$,
according to estimated flavor purity.
We determine directly from data
the average wrong-tag probabilities, 
$w_l \equiv (w_l^+ + w_l^-)/2~(l=1,6)$,
and differences between $\bz$ and $\bzb$ decays, 
$\dwl \equiv w_l^+ - w_l^-$,
where $w_l^{+(-)}$ is the wrong-tag probability
for the $\bz(\bzb)$ decay in each $r$ interval.

\subsection{Vertex reconstruction}
\label{sec:selection:vtx}

The vertex position for the $\fcp$ decay is reconstructed 
using leptons from $\jpsi$ decays or charged hadrons from $\eta_c$ decays.
Each vertex position is required to be consistent with
the interaction-region profile (IP), determined run-by-run,
smeared in the $r$-$\phi$ plane to account for the $B$ meson decay length.
With the IP constraint, we are able to determine a vertex
even with a single track;
the fraction of single-track vertices is about 10\% for $\zcp$.

The vertex position for the $\fflv$ decay that includes a $D$ meson in its 
decay products is reconstructed using the $D$ meson trajectory, a track other 
than the slow $\pi^-$ candidate from $D^{*-}$ decay, and the IP constraint.
For the $B^+ \to J/\psi K^+$ and $B^0\to\jpsi\kstarz(K^+\pi^-)$ decays, we
use leptons from $\jpsi$ decays and the IP constraint, in the same way as 
for the $\fcp$ decay.

The vertex position for the $\ftag$ 
is obtained with the IP constraint and with well reconstructed tracks
that are not assigned to $\fcp$ or $\fflv$. 
The algorithm is described
in detail elsewhere~\cite{bib:resol}.
The fraction of single-track vertices is about 22\% for $\ztag$.

We only use events with vertices that satisfy $|\Delta t| < 70$~ps
and $\xi<100$, where $\xi$ is the $\chi^2$ of the vertex fit 
calculated only in the $z$ direction~\cite{bib:resol}.
The overall vertex reconstruction efficiency is $87.1\pm0.7\%$ 
for $\bz\to\jpsi\ks$ candidates.

The proper-time interval resolution function $\Rsig(\Dt)$
is formed by convolving four components:
the detector resolutions for $\zcp$ and $\ztag$,
the shift in the $\ztag$ vertex position
due to secondary tracks originating from charmed particle decays,
and the kinematic approximation that the $B$ mesons are
at rest in the cms~\cite{bib:resol}.
A small component of broad outliers in the $\Dz$ distribution,
caused by mis-reconstruction, is represented by a Gaussian function.

\subsection{Signal yields}
\label{sec:selection:yields}
\begin{figure}
  \includegraphics[width=0.6\textwidth,clip]{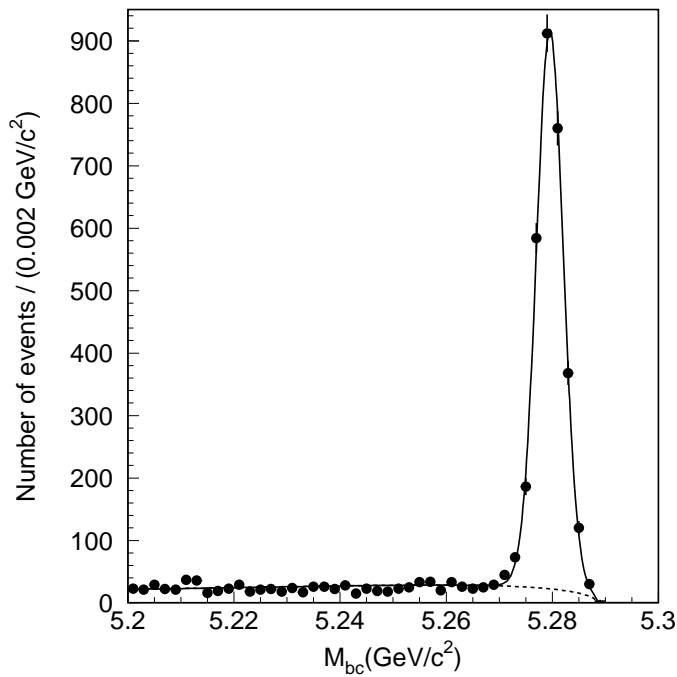}
  \caption{Beam-energy constrained mass distribution 
    within the $\dE$ signal region for all $f_{CP}$ modes other than $\jpsi\kl$. 
    The solid curve shows the fit to signal plus background distributions, and
    the dashed curve shows the background contribution.}
\label{fig:mbc} 
\end{figure}

\begin{figure}
  \includegraphics[width=0.6\textwidth,clip]{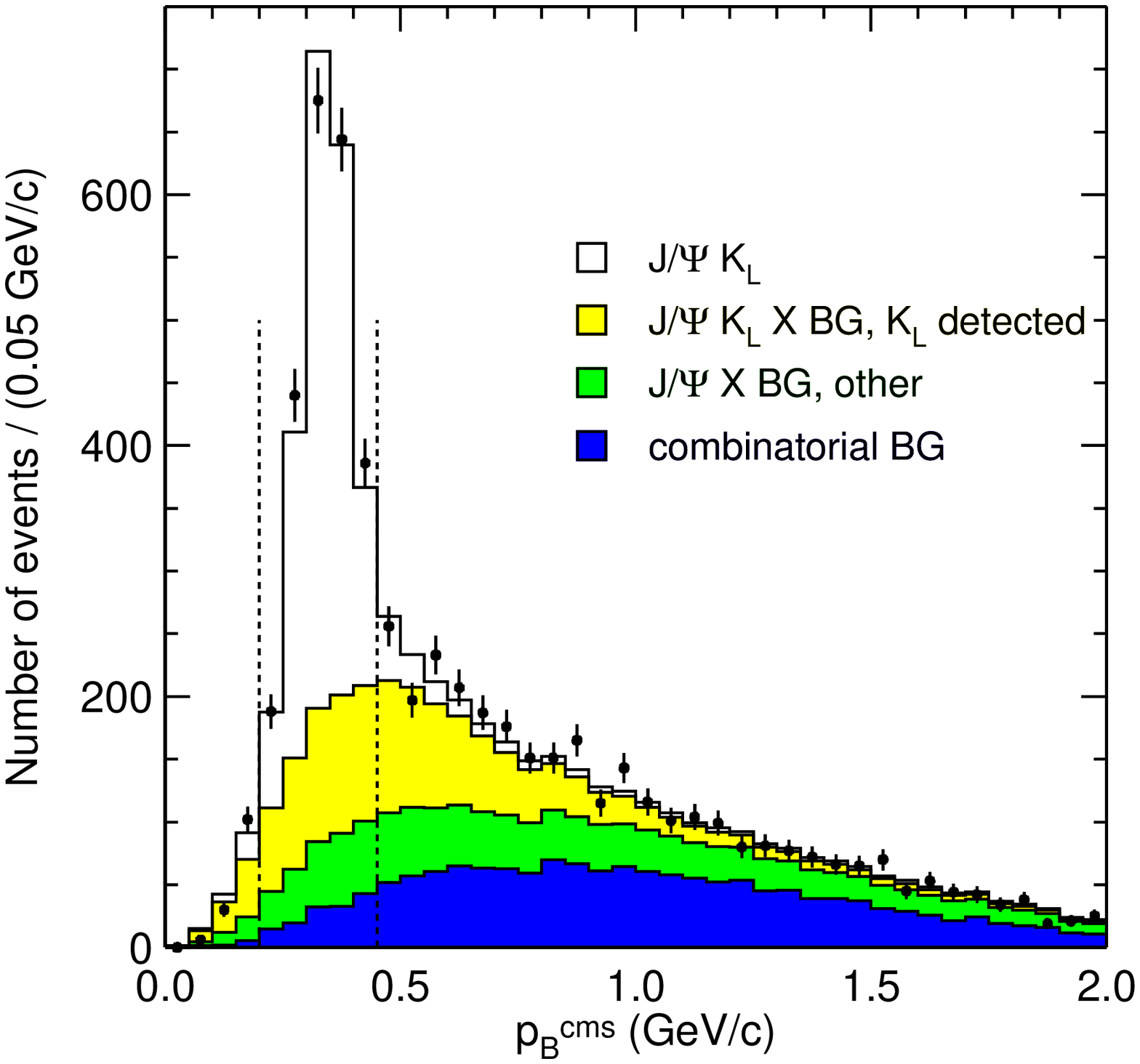}
  \caption{$p_B^\mathrm{cms}$ distribution for 
    $\bz \to \jpsi\kl$ candidates
    with the results of the fit. The dashed lines indicate the
    signal region.}
\label{fig:pbstar} 
\end{figure}

After flavor tagging and vertexing, we find 5417 $\fcp$ candidates
in total in the signal region; these are used for the $\sinbb$ determination.
Table~\ref{tab:number} lists the numbers of candidates, $\Nev$,
and the estimated signal purity for each $\fcp$ mode.
\begin{table}
  \caption{Numbers of reconstructed $B \to \fcp$
    candidates after flavor tagging and vertex reconstruction, $\Nev$,
    and the estimated signal purity, $p$.}
  \begin{ruledtabular}
    \begin{tabular}{llrl}
      \multicolumn{1}{c}{Mode} & $\xi_f$ & $\Nev$ & \multicolumn{1}{c}{$p$} \\
      \hline 
      $J/\psi \ks(\pip\pim)$       & $-1$ & 1997 & $0.976\pm 0.001$ \\
      $J/\psi \ks(\piz\piz)$       & $-1$ &  288 & $0.82~\pm 0.02$ \\
      $\psi(2S)(\ell^+\ell^-)\ks$  & $-1$ &  145 & $0.93~\pm 0.01$ \\
      $\psi(2S)(\jpsi\pip\pim)\ks$ & $-1$ &  163 & $0.88~\pm 0.01$ \\
      $\chi_{c1}(\jpsi\gamma)\ks$  & $-1$ &  101 & $0.92~\pm 0.01$ \\
      $\eta_c(\ks\km\pip)\ks$      & $-1$ &  123 & $0.72~\pm 0.03$ \\
      $\eta_c(\kp\km\piz)\ks$      & $-1$ &   74 & $0.70~\pm 0.04$ \\
      $\eta_c(p\overline{p})\ks$   & $-1$ &   20 & $0.91~\pm 0.02$ \\
      \cline{3-4}
      All with $\xi_f = -1$        & $-1$ & 2911 & $0.933\pm 0.002$ \\
      \hline
      $J/\psi\kstarz(\ks\piz)$ & +1(81\%)
                                          &  174 & $0.93~\pm 0.01$ \\
      \hline
      $J/\psi\kl$                  & $+1$ & 2332 & $0.63~\pm 0.03$ \\
    \end{tabular}
  \end{ruledtabular}
\label{tab:number} 
\end{table}
%
%
Figure~\ref{fig:mbc} shows the $\mb$ distribution
after applying mode-dependent requirements on $\dE$
for all $\bz$ candidates except for $\bz \to \jpsi\kl$.
There are 3085 entries in total in the signal region defined
as $5.27~\text{GeV/}c^2 < \mb < 5.29$ GeV/$c^2$.
Figure~\ref{fig:pbstar} shows the $\pB$ distribution
for $\bz \to \jpsi\kl$ candidates. We find 2332 entries
in the $0.20~\text{GeV/}c \le \pB \le 0.45$ GeV/$c$ signal region.
%
%

\begin{figure}[htbp]
  \begin{center}
    \includegraphics[width=0.8\textwidth,clip]{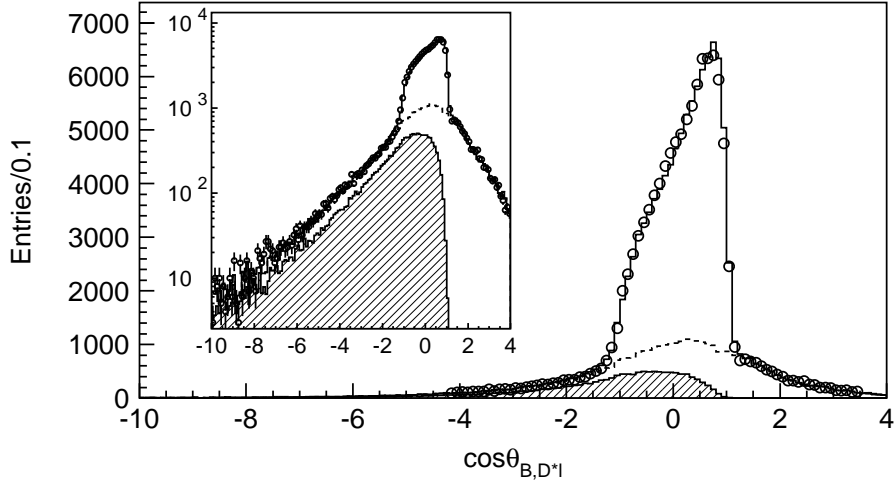}
  \end{center}
 \caption{$\cosbdl$ distribution for the $\dslnu$ candidates. 
The circles with errors show
the data. The solid line is the fit result. The total background
and the $D^{**}\ell\nu$ component are shown by the dashed line
and the hatched area, respectively. The inset shows the same figure with
a logarithmic vertical scale.} 
\label{fig:cosb}
\end{figure}

\begin{figure}[htbp]
  \begin{center}
    \includegraphics[width=0.45\textwidth,clip]{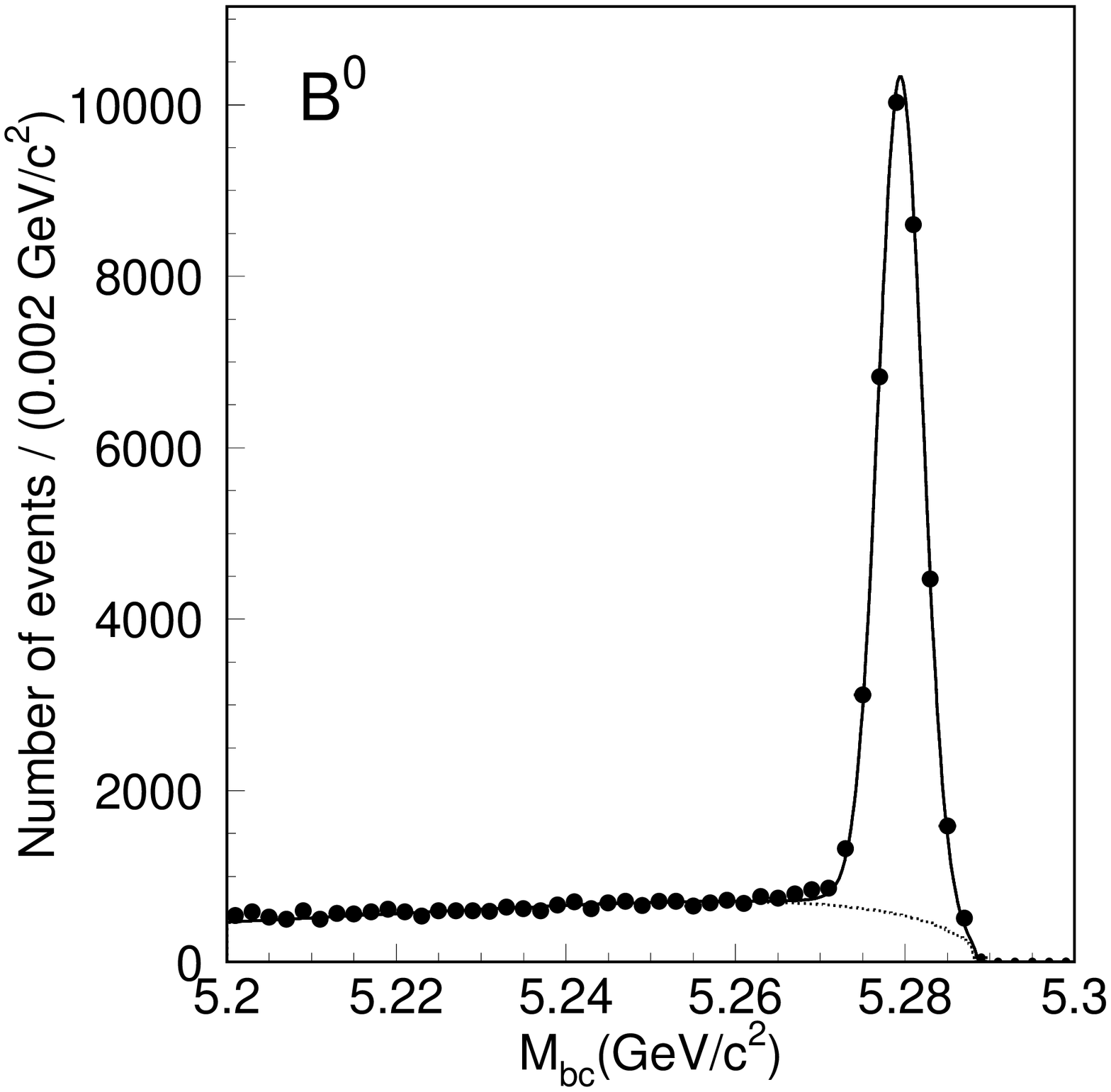}
    \includegraphics[width=0.45\textwidth,clip]{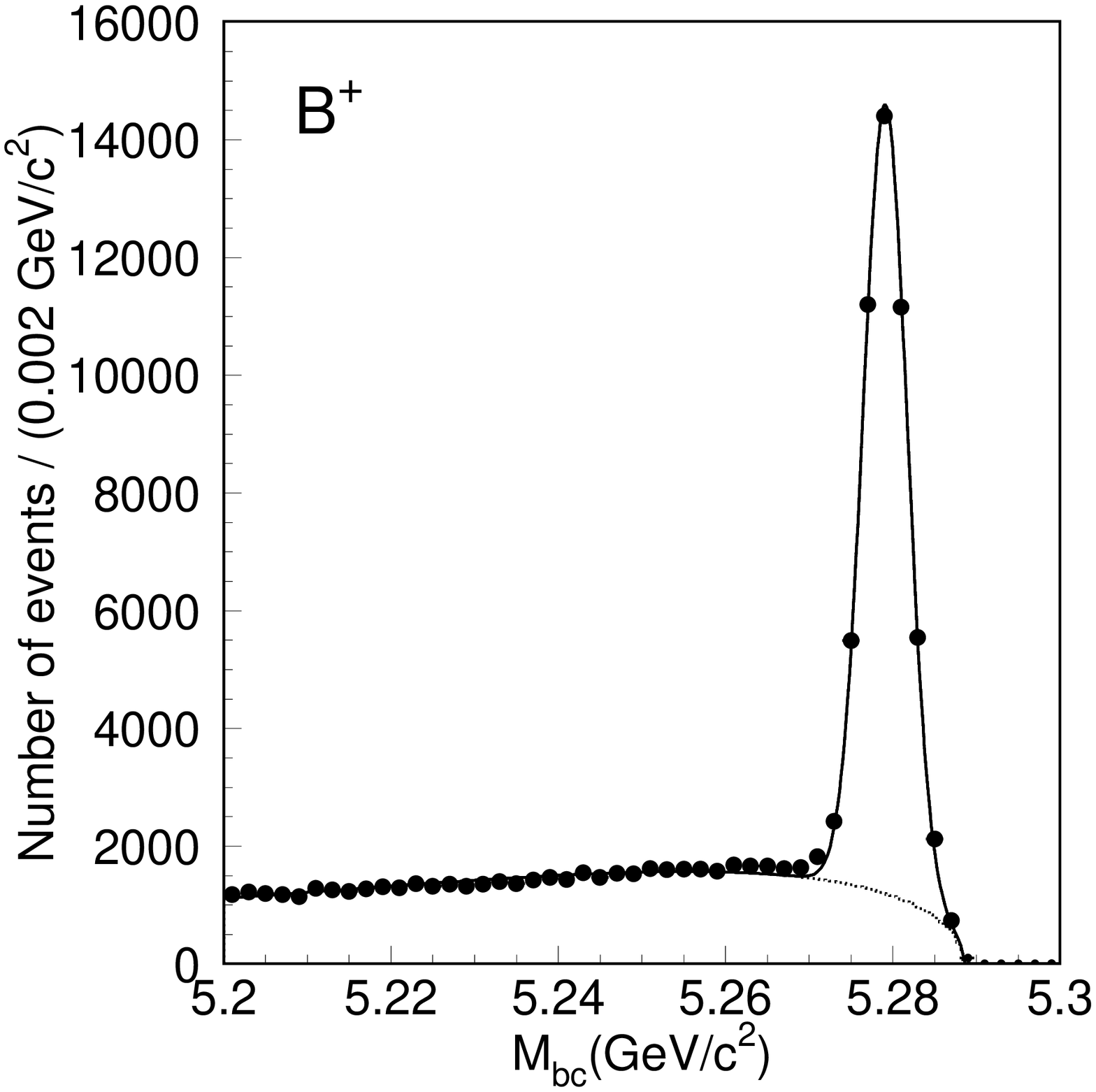}
  \end{center}
 \caption{Beam-energy constrained mass distributions 
    in the $\dE$ signal region
    for flavor-eigenstate candidates in neutral $B$ meson decays (left)
    and in charged $B$ meson decays (right).
    The solid curves show the fits to signal plus background distributions and
    the dashed curves show the background contributions.}
\label{fig:fflv}
\end{figure}

Table~\ref{tab:yield:cntrl} lists $\Nev$ and the purity for
each $\fflv$ mode after the vertexing.
The total number of $\fflv$ candidates is
177368 with a purity of 81\%. Figure~\ref{fig:cosb} shows
the $\cosbdl$ distribution for the $\dslnu$ candidates. 
Figure~\ref{fig:fflv} shows the 
$\mb$ distributions for $\bz$ and $\bp$ decays to
$\fflv$ states.

According to a MC simulation study, there is a small fraction 
of background
(less than $1$\% for $\fcp$ and $3$\% for $\fflv$ candidates)
from other $B$ decays peaking in the $\mb$ signal region.
The effect of the peaking background is treated as a systematic error.
%
\begin{table}
  \caption{Numbers of reconstructed $B \to \fflv$ candidates 
    after vertex reconstruction, $\Nev$,
    and the estimated signal purity, $p$.
    $J/\psi\ks$ candidates are used with no flavor assignment.}
  \begin{ruledtabular}
    \begin{tabular}{lrl}
      \multicolumn{1}{c}{Mode} & $\Nev$ & \multicolumn{1}{c}{$p$} \\
      \hline
      $\dslnu$         & 84823  & 0.781 \\
      $\dsm\pip$       & 11921  & 0.888 \\
      $\dm\pip$        & 11156  & 0.899  \\
      $\dsm\rhop$      &  8767  & 0.763 \\
      $\jpsi\kstarz(\kp\pim)$
                       &  3681  & 0.954 \\
      $\jpsi\ks(\pip\pim)$
                       &  2001  & 0.976 \\
      \hline
      $\bz$ total      & 122349 & 0.809 \\
      \hline \hline
      $\dzb\pip$       & 46248 & 0.783 \\
      $\jpsi\kp$       &  8771 & 0.966 \\
      \hline
      $\bp$ total      & 55019 & 0.812 \\
      \hline \hline
      $\bz + \bp$ total & 177368 & 0.810 \\
    \end{tabular}
  \end{ruledtabular}
\label{tab:yield:cntrl}
\end{table}
%

\section{Fit results with flavor-eigenstate samples}
\label{sec:control}
We perform a multi-parameter fit
to flavor-eigenstate samples
to obtain the $B$ meson lifetimes,
the $\bz$-$\bzb$ mixing parameter $\dM$,
wrong-tag fractions, and parameters for the resolution function simultaneously.
We use $\bz \to D^{*-}\ell^+\nu$, $\jpsi\kstarz(\kp\pim)$,
$D^{*-}\pi^+$, $D^-\pi^+$, $D^{*-}\rho^+$, and
$J/\psi \ks(\ell^+\ell^-)$ (for $\taubz$ and resolution parameters only)
for $\bz$ decays,
and $B^+ \to \overline{D}{}^0\pi^+$ and $J/\psi K^+$
for $B^+$ decays.
The fit uses 32 parameters; 12 for wrong-tag fractions,
14 for the resolution function,
3 for the $B^+$ background in $\bz$ decays, and
3 physics parameters $\dM$, $\taubz$ and $\taubp$.
We also obtain the lifetime ratio, $r_{\tau_{B}}\equiv \taubp/\taubz$, 
by repeating the fit in which $\taubp$ is replaced with $r_{\tau_{B}}\taubz$.
Two of the 14 parameters for the resolution function 
are newly added to the resolution function
described in~\cite{bib:resol}
to improve the description of the
effect of charmed particle decays on the $\ftag$ vertex.

The probability density function (PDF) expected
for the signal distribution for $\bz$ decays to $\fflv$ is given by
\begin{equation}
    {\cal P}^{\rm OF[SF]}_{\rm mix}(\dt, q, w_l, \dwl) 
    = \frac{e^{-|\dt|/\taub}}{8\taub}
    \Bigl\{1 -q\dwl \nonumber
    +[-](1-2w_l)\cos (\dmd \dt)\Bigr\},
\end{equation}
where OF (SF) denotes $\bz\bzb$ ($\bz\bz$ or $\bzb\bzb$),
i.e. a state with the opposite (same) flavor.
The signal PDF for $\bp$ decays is given by
\begin{equation}
\label{eq:bpluspdf}
    {\cal P}^{B^+}_{\mathrm{sig}}(\dt)
    = \frac{e^{-|\dt|/\taubp}}{2\taubp}.
\end{equation}
The signal PDFs are convolved with the $\Rsig(\Dt)$ to account for the
detector resolution.

The background PDF for the hadronic modes is modeled as a sum of exponential
and prompt components,
\begin{equation}
\label{eq:bkgpdf}
\Pbkg(\Delta t) =
(1-f_{\delta})
\frac{e^{-|\Delta t|/\tau_{\mathrm{bkg}}}}
{2\tau_{\mathrm{bkg}}}
+f_{\delta}\delta (\Delta t),
\end{equation}
($1- f_{\delta}$) is the fraction of the exponential component with the
effective lifetime $\tau_{\mathrm{bkg}}$,
and $\delta(\Delta t)$ is the Dirac delta function.
It is convolved with a sum of two Gaussians, 
which is used as the background resolution function.
The parameters for the background PDF are determined using the
$\Delta E$-$\mb$ sideband region for each decay mode.
For $B^+ \to \dzb\pip$ decays, 
using events outside the signal region,
the value for $f_{\delta}$ is determined to be
$0.49\pm0.01$ ($0.45 \pm 0.03$) for events with 
multi-track (single-track) vertices and
the effective lifetime $\tau_{\mathrm{bkg}}$ 
is found to be $0.93\pm0.03$~ps.
The parameters for other $B\to \fflv$ decays that include a
$D$ meson as a decay product
are similar to those for the
$B^+ \to \dzb\pip$ decay.
A similar procedure for $B^+ \to \jpsi\kp$ decays yields
$f_{\delta} = 0.86\pm0.05$ ($0.75\pm0.08$)
for events with multi-track (single-track) vertices and
$\tau_{\mathrm{bkg}} = 1.55\pm0.22$~ps.
The parameters for $\bz\to\jpsi\ks$ are 
similar to those for $B^+ \to \jpsi\kp$.
The value for $f_{\delta}$ in $\bz\to\jpsi\kstarz(\kp\pim)$ decays 
is found to be small;
$f_{\delta}$ is $0.07\pm0.05$ for events with multi-track vertices and
is fixed at zero for events with single-track vertices.
The effective lifetime $\tau_{\mathrm{bkg}}$ is $1.50\pm0.05$~ps.


The background for the $\dslnu$ decay is divided into four components:
$B \to D^{**}\ell\nu$ events ($8.7\pm0.3$\%);
fake $D^*$ mesons ($8.0\pm0.1$\%);
random combination of $D^*$ mesons with leptons with no angular correlation
called ``uncorrelated background'' ($2.5\pm0.1$\%);
continuum events ($2.7\pm0.2$\%).
Here $D^{**}$ consists of
charmed mesons heavier than the $D^*$ meson and
non-resonant $D^*\pi$ components. 
The PDF for the $B\to D^{**}\ell\nu$ background is given by a sum of $B^0$ 
and $B^+$ components,
\begin{equation}
\mathcal{P}_{D^{**}\ell\nu}^{OF[SF]}(\Delta t) = 
(1-f_{B^+})\mathcal{P}_{\rm mix}^{OF[SF]}+f_{B^+}\mathcal{P}_{B^+}^{OF[SF]},
\end{equation}
where $f_{B^+}$ is the $B^+$ fraction in the $B\to D^{**}\ell\nu$ background.
The $\mathcal{P}_{B^+}^{\rm OF[SF]}$ is given by
$\mathcal{P}_{B^+}^{\rm OF}(\Delta t)=(1-w^l_{B^+})P^{B+}_{\rm bkg}(\Delta t)$
and
$\mathcal{P}_{B^+}^{\rm SF}(\Delta t)=w^l_{B^+}P^{B+}_{\rm bkg}(\Delta t)$, 
where $w^l_{B^+}$ is the wrong tag fraction determined from 
the $B^+ \to \dzb\pip$ sample and $\mathcal{P}_{\rm bkg}^{B^+}$ is given by
\begin{equation}
\mathcal{P}_{\rm bkg}^{B^+}(\Delta t)
    = (1-f_{\taubppr})\frac{e^{-|\dt|/\taubp}}{4\taubp}
    + f_{\taubppr}\frac{e^{-|\dt|/\taubppr}}{4\taubppr}.
\end{equation}
Here $f_{\taubppr}$ and $\taubppr$ are
the fraction and the effective lifetime for events in which
an additional $\pip$ from the $D^{**}$ decays
contaminates the $\ftag$ vertex reconstruction.
The parameters $f_{B^+}$, $f_{\taubppr}$ and $\taubppr$ are determined in the 
final fit.
To determine these parameters precisely, events in  $-10<\cosbdl<-1.1$, where 
the $D^{**}\ell\nu$ background events are dominant, 
are also included in the fit.
The fit yields $f_{B^+}=0.51\pm0.04$, 
$f_{\taubppr}=0.56\pm0.10$ and $\taubppr=0.74\pm0.14$~ps.

For continuum and uncorrelated backgrounds, the same functional form as that of 
the hadronic background PDF is used.
The parameters for continuum are determined from
off-resonance data to be $f_{\delta} = 0.55\pm0.09$ ($0.58\pm0.11$)
for events with multi-track (single-track) vertices and
$\tau_{\mathrm{bkg}} = 0.80\pm0.08$~ps.
For the uncorrelated background, a fit to the sample outside the signal region
yields $f_{\delta}=0.15\pm0.08$ and $\tau_{\mathrm{bkg}} = 1.23\pm0.06$~ps.

The PDF of the fake $D^*$ background is given by Eq.(\ref{eq:bkgpdf})
with a mixing component added to account for oscillation in the background.
A fit to events in the $\mdiff$ sideband yields 
$f_{\delta} = 0.13\pm0.02$ ($0.04\pm0.03$) for events 
with multi-track (single-track) vertices,
$\tau_{\mathrm{bkg}} = 1.49\pm0.03$~ps and
$\Delta m_{\mathrm bkg}=0.54\pm0.05$~ps$^{-1}$, where 
$\Delta m_{\mathrm bkg}$ is the effective mixing parameter.
The fraction of the mixing component and wrong tag fractions are 
determined for each of the six intervals of the flavor tag quality $r$.
The wrong tag fractions range from $0.50\pm0.01$ for the lowest $r$ region
to $0.20\pm0.01$ for the highest $r$ region.
The fraction of the mixing component for the lowest $r$ region is fixed at 0.
Values 
for the other $r$ intervals range from $0.39\pm0.12$ to $0.84\pm0.09$.

We test the fit method and parameterization with
a large number of MC events, and
obtain results consistent with the input values.
The wrong-tag fractions obtained 
with the MC events are also found to be correct.

The unbinned maximum-likelihood fit to data yields
\begin{eqnarray}
\taubz     &=& \taubzresult, \\
\taubp     &=& \taubpresult,\\
\taubratio &=& \taubratioresult,\\
\dM        &=& \dmdresult.
\end{eqnarray}
The results are consistent with our previous 
measurements~\cite{Abe:2002id,Tomura:2002qs,Hara:2002mq}
and supersede them.
\begin{figure}
  \includegraphics[width=0.6\textwidth,clip]{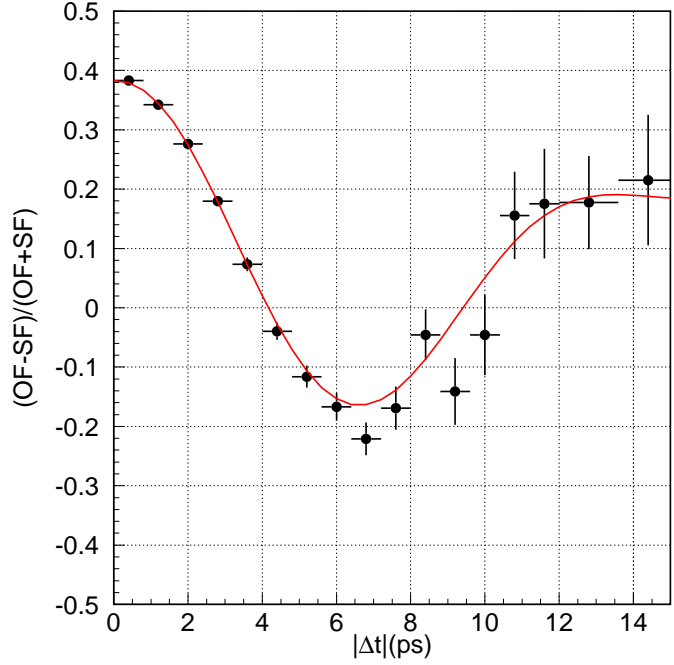}
  \caption{Time-dependent flavor asymmetry
  for flavor-eigenstate decays. The
  curve is the result of the unbinned maximum-likelihood fit.}
  \label{fig:dmd}
\end{figure}
Figure~\ref{fig:dmd} shows the 
flavor asymmetry,
${\cal A}(\dt)=[{\rm N}_{\rm OF}(\dt)-{\rm N}_{\rm SF}(\dt)]/[{\rm N}_{\rm OF}(\dt)+{\rm N}_{\rm SF}(\dt)]$,
where ${\rm N}_{\rm OF(SF)}$ denotes the number of OF (SF) events.
%
\begin{figure}
  \includegraphics[width=0.45\textwidth,clip]{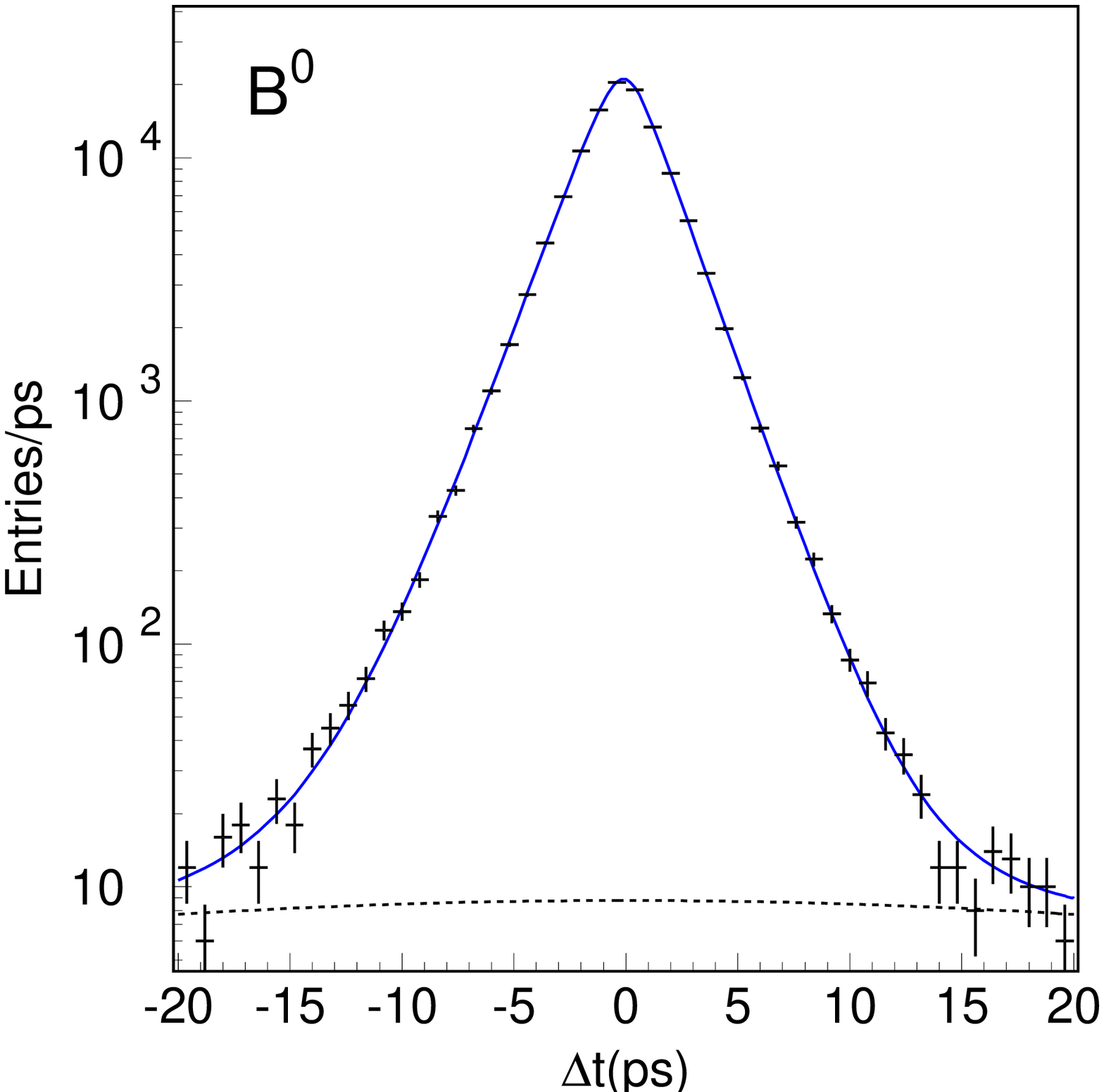}
  \includegraphics[width=0.45\textwidth,clip]{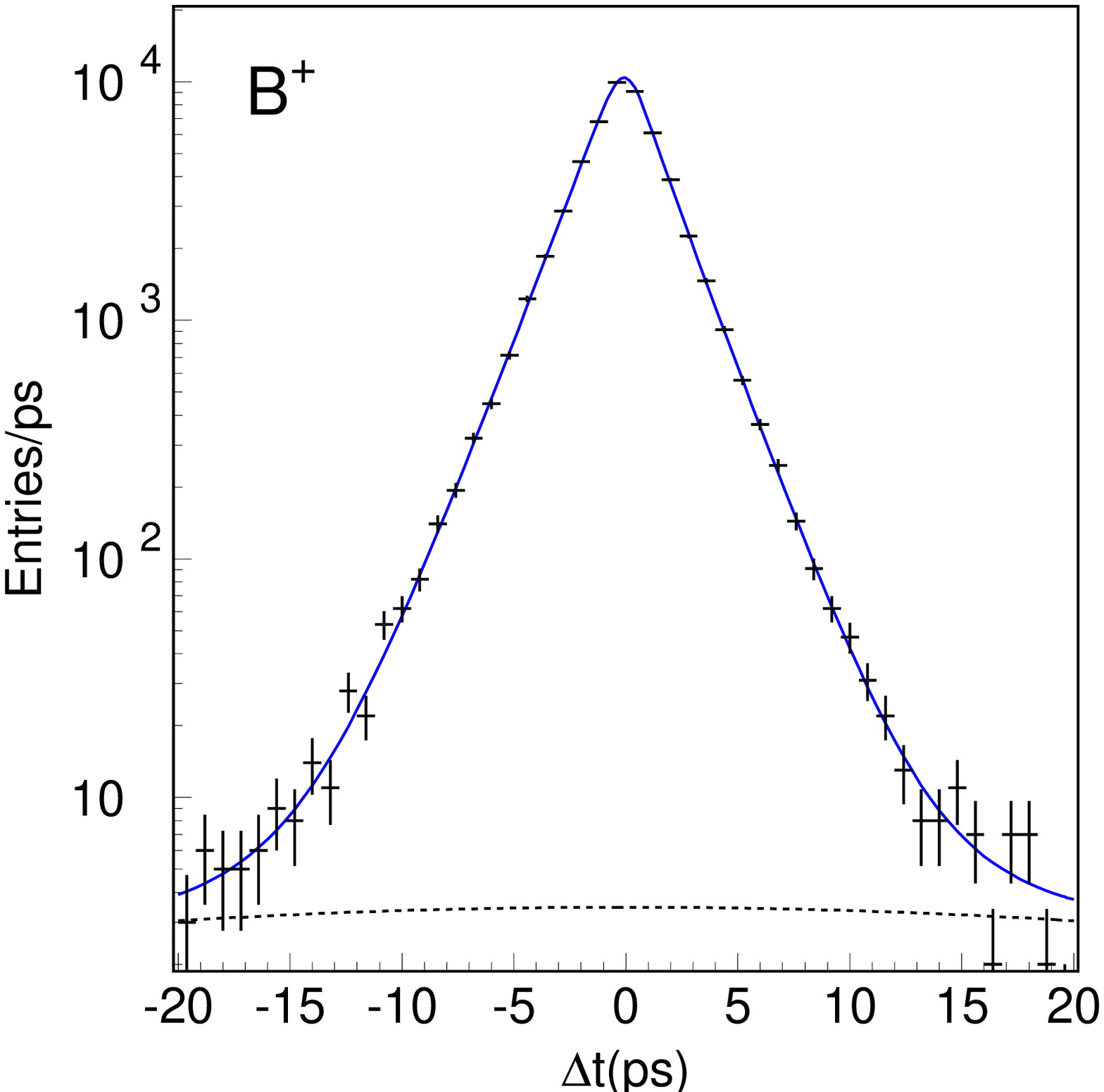}
  \caption{$\Dt$ distributions of neutral $B$ meson pairs (left)
  and charged $B$ meson pairs (right).
The solid lines represent
the results of the unbinned maximum-likelihood fit.
The dashed lines correspond to the outlier components.}
  \label{fig:blife}
\end{figure}
The results of the lifetime measurements
for neutral and charged $B$ meson decays
are shown in Fig.~\ref{fig:blife}.

\begin{table}[hbtp]
  \caption{Summary of the systematic errors on the measurement of
$\taubz$, $\taubp$, $\taubratio$ and $\dmd$.}
  \begin{ruledtabular}
  \begin{tabular}{lrrrr}
   Source & $\taubz$ & $\taubp$  & $\taubratio$ & $\dmd$\\ \hline 
   Vertex reconstruction         & 0.005 & 0.007 & 0.003    & 0.003\\
   Resolution function           & 0.004 & 0.005 & 0.004    & 0.001\\
   Possible fit bias             & 0.003 & 0.004 & 0.003    & 0.002\\
   $D^{**}\ell\nu$ background    & 0.004 & 0.003 & 0.002    & 0.004\\
   Other background fraction     & 0.001 & 0.006 & 0.003    & 0.001\\
   Background $\Dt$ shape        & 0.006 & 0.003 & 0.005    & 0.002\\
\hline
   Total                         & 0.010 & 0.011 & 0.008    & 0.006\\
                                 & (ps)  & (ps)  &          & (ps$^{-1}$)\\
  \end{tabular}
  \end{ruledtabular}
  \label{tab:syserr:cntrl}
\end{table}
Systematic uncertainties are listed in Table~\ref{tab:syserr:cntrl}.
The method to determine the systematic errors due to
the vertex reconstruction follows the same procedure
as for the $\sinbb$ measurement, which will be explained
later. We estimate the contribution
due to uncertainties in the resolution function
by comparison of different parameterizations, as well as
by changing parameters that are derived from MC to
model the effect of non-primary tracks~\cite{bib:resol}.
A possible bias in the event reconstruction and fitting procedure 
is checked with a large number of MC events.
We find no bias and take the statistical error in MC as a systematic error.
Several $D^{**}$ components are used in this analysis
to model the $\cosbdl$ shape for the $D^{**}\ell\nu$ background.
To estimate the systematic errors due to uncertainties of 
the fractions of the $D^{**}$ components,
we set the fraction of each component to unity
(with all other components set to zero) and repeat the analysis;
for each measurement,
we take the largest variation on the result
as the systematic error.
Systematic errors that arise from uncertainties in other background fractions 
and from the background $\Dt$ shape are obtained by varying
each parameter individually, repeating the fit procedure,
and adding each contribution in quadrature.
In the nominal fit, we do not include a mixing component in the background 
PDF for the hadronic decays. We repeat the fit with a background PDF
including a mixing term.
Uncertainties in the overall $z$ scale of the detector 
arising from the measurement error and thermal expansion
during the operation are found to be negligible.

The same fit also yields wrong-tag fractions that
are summarized in Table~\ref{tab:wtag}.
\begin{table}
  \caption{Event fractions $\epsilon_l$,
    wrong-tag fractions $w_l$, wrong-tag fraction differences $\dwl$,
    and average effective tagging efficiencies
    $\eeff^l = \epsilon_l(1-2w_l)^2$ for each $r$ interval.
    The errors include both statistical and systematic uncertainties.}
  \begin{ruledtabular}
    \begin{tabular}{ccclll}
      $l$ & $r$ interval & $\epsilon_l$ &\multicolumn{1}{c}{$w_l$} 
          & \multicolumn{1}{c}{$\dwl$}  &\multicolumn{1}{c}{$\eeff^l$} \\
      \hline
 1 & 0.000 -- 0.250 & 0.398 & $0.464\pm0.006$ &$-0.011\pm0.006$ &$0.002\pm0.001$ \\
 2 & 0.250 -- 0.500 & 0.146 & $0.331\pm0.008$ &$+0.004\pm0.010$ &$0.017\pm0.002$ \\
 3 & 0.500 -- 0.625 & 0.104 & $0.231\pm0.009$ &$-0.011\pm0.010$ &$0.030\pm0.002$ \\
 4 & 0.625 -- 0.750 & 0.122 & $0.163\pm0.008$ &$-0.007\pm0.009$ &$0.055\pm0.003$ \\
 5 & 0.750 -- 0.875 & 0.094 & $0.109\pm0.007$ &$+0.016\pm0.009$ &$0.057\pm0.002$ \\
 6 & 0.875 -- 1.000 & 0.136 & $0.020\pm0.005$ &$+0.003\pm0.006$ &$0.126\pm0.003$ \\
    \end{tabular}
  \end{ruledtabular}
\label{tab:wtag} 
\end{table}
The total effective tagging efficiency is determined to be
$\eeff \equiv \sum_{l=1}^6 \epsilon_l(1-2w_l)^2 = \efftot$,
where $\epsilon_l$ is the event fraction for each $r$ interval
determined from the $\jpsi\ks$ simulation and is listed in
Table~\ref{tab:wtag}.
The error includes both statistical and systematic uncertainties.

We find that the average $\Dt$ resolution is $\sim 1.43$~ps (rms).
The width of the outlier component
is determined to be $(39\pm 2)$~ps;
the fractions of the outlier components are $(2.1 \pm 0.6) \times 10^{-4}$
for events with both vertices reconstructed with more than one track,
and $(3.1 \pm 0.1) \times 10^{-2}$ for events with at least
one single-track vertex.

\section{Results of {\boldmath $CP$} asymmetry measurements}

\begin{figure}
  \includegraphics[width=0.6\textwidth,clip]{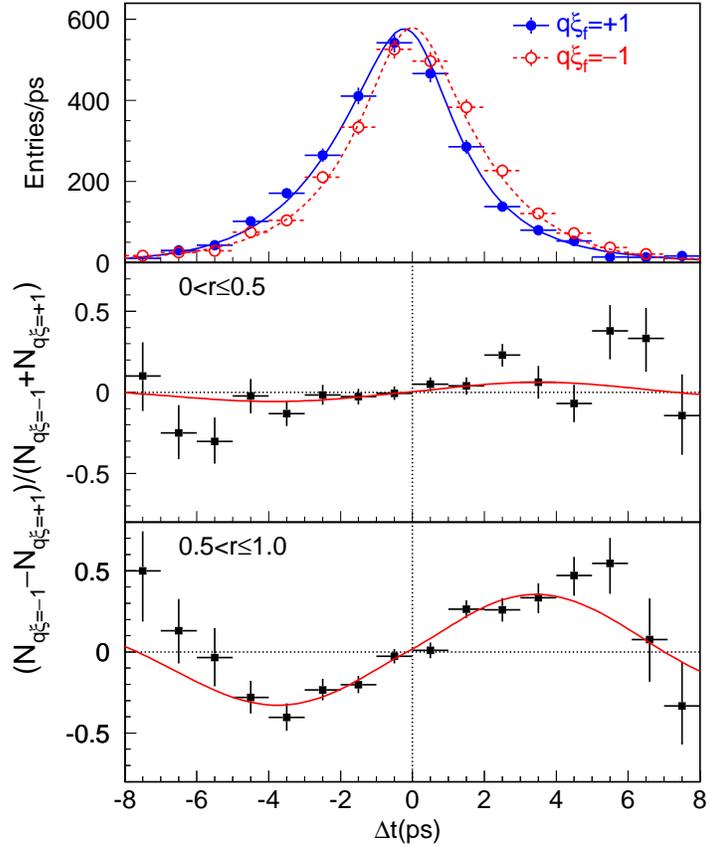}
  \caption{$\Dt$ distributions for the events
    with $q\xi_f = -1$ (open points) and 
    $q\xi_f = +1$ (solid points) with all modes combined (top),
    asymmetry between $q\xi_f=-1$ and $q\xi_f=+1$
    samples with $0 < r \le 0.5$ (middle), and
    with $0.5 < r \le 1$ (bottom).
    The results of the global unbinned maximum-likelihood fit 
    ($\sinbb=\sinbbcenter$) are also shown.}
\label{fig:cpfit} 
\end{figure}

Figure~\ref{fig:cpfit} shows the observed $\Dt$ distributions
for the $q\xi_f = +1$ and $q\xi_f = -1$ event samples (top), 
the asymmetry between two samples with $0 < r \le 0.5$ (middle)
and with $0.5 < r \le 1.0$ (bottom).
The asymmetry in the region $0.5 < r \le 1.0$,
where wrong-tag fractions are small
as shown in Table~\ref{tab:wtag},
clearly demonstrates large $CP$ violation.

We determine $\sinbb$ from an unbinned maximum-likelihood fit
to the observed $\Dt$ distributions.
The PDF for the signal distribution is given by
\begin{equation}
  \label{eq:deltat}
  \Psig(\Dt, q, w_l, \dwl, \xi_f) =
  \frac{e^{-|\Dt|/\taubz}}{4\taubz}
  \Bigl[ 1 - q\dwl
  -q\xi_f(1-2w_l)\sinbb\sin(\dM\Dt) \Bigr],
\end{equation}
where we fix the $\bz$ lifetime $\taubz$ and mass difference $\dM$
at their world average values~\cite{bib:PDG2004}.
Each PDF is convolved with the appropriate $\Rsig(\Dt)$
to determine the likelihood value for each event as a function of $\sinbb$:
\begin{eqnarray}
  P_i &=& (1-\fol) \int_{-\infty}^{+\infty} \Bigl[ \fsig \Psig(\Dt')\Rsig(\Dt-\Dt')
  \nonumber \\
  &+& (1-\fsig)\Pbkg(\Dt')\Rbkg(\Dt-\Dt')\Bigr] d\Dt'
  + \fol \Pol(\Dt),
\end{eqnarray}
where $\fsig$ is the signal fraction
calculated as a function of $\pB$ for $\jpsi\kl$
and of $\dE$ and $\mb$ for other modes.
$\Pbkg(\Dt)$ is the PDF for combinatorial background events,
which is modeled as a sum of exponential and prompt components.
It is convolved with a sum of two Gaussians, $\Rbkg$,
which is used as the background resolution function.
We assume no asymmetry in the background $\Delta t$ distribution.

For $\bz\to\jpsi\kl$ and $\jpsi\kstarz$ decays, in addition to the 
combinatorial background, background events from other $B$ decays and their
$CP$ asymmetries are considered.
The background in the $\jpsi\kl$ mode is dominated by
the following $B\to\jpsi X$ decays:
$\jpsi\ks$ having $\xi_f=-1$ ($10\pm2$\%);
$\psi(2S)\kl$, $\chi_{c1}\kl$ and $\jpsi \piz$ having $\xi_f=+1$ ($4\pm1$\%);
$\jpsi\kstarz(\kstarz\to\kl\piz)$ ($20\pm2$\%), which is a mixture of $\xi_f=-1$ (81\%)
and $\xi_f=+1$ (19\%);
other non-$CP$ modes ($66\pm1$\%).
The fraction of each component is obtained from a MC simulation study.
For $\jpsi\kstarz(\ks\piz)$ decays, 
we include in our PDF 
contaminations from other $B\to\jpsi\kstar$
decays ($7.1\pm0.3$\%)
and non-resonant $\bz\to\jpsi\ks\piz$ decays ($6.3\pm0.5$\%) 
in the $\mb$ peak.
The background fractions are 
obtained from MC and the $\kstar$ mass sideband~\cite{bib:itoh}.
We use the signal PDF with no $CP$ asymmetry for these components.

To account for a small number of events that give large $\Dt$ in
both the signal and background, we introduce
the PDF of the outlier component, $\Pol$, and its fraction $\fol$.
The only free parameter in the final fit is $\sinbb$,
which is determined by maximizing the likelihood function $L = \prod_i P_i$,
where the product is over all events.
We obtain
\begin{equation}
\sinbb = \sinbbresult .
\end{equation}
The result is consistent with the value in our previous 
publication~\cite{bib:Belle_sin2phi1_78fb-1} 
and supersedes it with a reduced error.

The signal PDF for a neutral $B$ meson decaying into a $CP$ eigenstate
[Eq.~(\ref{eq:deltat})] can be expressed in a more general form as
\begin{eqnarray}
  \label{eq:deltat_general}
  \Psig(\Dt, q, w_l,\dwl)
  &=& \frac{ e^{-|\Dt|/\taubz} }{4\taubz}
  \Biggl\{ 1 -q\dwl \nonumber \\
  &+& q(1-2w_l)
  \Bigl[ \cals \sin(\dM\Dt) 
  + \cala \cos(\dM\Dt) \Bigr] \Biggr\},
\end{eqnarray}
where $\cals \equiv 2{\rm Im}(\lambda)/(|\lambda|^2+1)$,
$\cala \equiv (|\lambda|^2-1)/(|\lambda|^2+1)$,
and $\lambda$ is a complex parameter that depends on both
$\bz$-$\bzb$ mixing and on the amplitudes for $\bz$ and $\bzb$ decay
to a $CP$ eigenstate.
The presence of the cosine term ($|\lambda| \neq 1$)
would indicate direct $CP$ violation;
the value for $\sinbb$ reported above is determined
with the assumption $|\lambda| = 1$, as $|\lambda|$ is expected
to be very close to one in the SM.
In order to test this assumption,
we also performed a fit using the expression above with
$a_{CP} \equiv -\xi_f {\rm Im}(\lambda)/|\lambda|$
and $|\lambda|$ as free parameters, keeping everything else the same.
We obtain
\begin{equation}
|\lambda| = \lambdaresult,
\end{equation}
and
$a_{CP} = \sinbbcenter\pm\sinbbstat\mathrm{(stat)}$.
This result is consistent with the assumption of
no direct $CP$ violation used in our analysis and
the $a_{CP}$ term is in good agreement with the $\sinbb$ value obtained
with the one-parameter fit.

\begin{table}[hbtp]
  \caption{Summary of the systematic errors on $\sinbb$ and $|\lambda|$.}
  \begin{ruledtabular}
  \begin{tabular}{lrr}
   Source                                    & $\sinbb$ & $|\lambda|$\\ 
  \hline 
   Vertex reconstruction                        & 0.013 & 0.012\\
   Flavor tagging                               & 0.007 & 0.008\\
   Resolution function                          & 0.008 & 0.004\\
   Possible fit bias                            & 0.008 & 0.006\\
   Background fraction ($J/\psi\kl$)            & 0.011 & 0.003\\
   Background fraction (except for $J/\psi\kl$) & 0.007 & 0.007\\
   Physics ($\taubz$, $\dmd$, $J/\psi\kstarz$)  & 0.003 & 0.001\\
   Background $\Dt$ shape                       & 0.002 & 0.001\\
   Tag-side interference                        & 0.002 & 0.028\\
\hline
   Total                                        & 0.023 & 0.033\\
  \end{tabular} 
  \end{ruledtabular}
  \label{tab:syserr:cp}
\end{table}
Table~\ref{tab:syserr:cp} lists the systematic errors
on $\sinbb$ and $|\lambda|$.
The total systematic error is obtained
by adding each of them in quadrature.
The largest contribution for $\sinbb$
comes from vertex reconstruction.
The systematic error due to the IP constraint 
in the vertex reconstruction is estimated by
varying ($\pm10~\mu$m) the smearing used to account for the
$B$ flight length.
The track selection criteria
are also varied to search for possible systematic biases.
The effect of the vertex quality cut is estimated by varying
the cut to $\xi<50$ and $\xi<200$.
We vary the $|\Delta t|$ range by $\pm 30$~ps
to estimate the systematic uncertainty due to the $|\Delta t|$
fit range.
Small biases in the $\Dz$ measurement 
are observed in $e^+e^-\to\mu^+\mu^-$ and other control
samples. Systematic errors 
are estimated by applying special correction functions
to account for the observed biases, repeating
the fit, and comparing the obtained values with the nominal results.
Systematic errors due to imperfect SVD alignment
are determined
from MC samples that have artificial mis-alignment effects
to reproduce impact-parameter resolutions observed in data.
In these studies, whenever required, 
we repeat the fit to the $\fflv$ samples,
update resolution function parameters and wrong tag fractions,
and perform the fit to $CP$-eigenstate event samples
using the updated parameters so that the uncertainties in question
are treated in a consistent way.

Systematic errors due to uncertainties in the wrong tag
fractions given in Table~\ref{tab:wtag} are studied by varying
the wrong tag fraction individually for each $r$ region.
Possible differences of the tagging performance
between $\fcp$ and $\fflv$ events are estimated using MC events.

Systematic errors due to uncertainties in the resolution function
are estimated by varying each resolution parameter obtained from
data (MC) by $\pm 1\sigma$ ($\pm 2\sigma$), repeating the fit
and adding each variation in quadrature. We also
divide the entire data set into two and prepare two sets of
resolution parameters to consider a possible difference in
the detector performance. 
We repeat the fit with these resolution parameters and
assign the difference from the nominal result as a systematic error.
We also include other sources
examined for the fit to the flavor-eigenstate samples,
which are explained in the previous section.

A possible fit bias is examined by a fit to a large number of MC events.
We find no bias and 
take the statistical error from the MC as a systematic error.

Systematic errors from uncertainties in the background fractions
and in the background $\Dt$ shape
are estimated by varying each background parameter obtained
from data (MC) by $\pm 1\sigma$ ($\pm 2\sigma$).
The systematic error due to $CP$ content in the $\jpsi\kl$
backgrounds is checked by varying the parameters obtained from the MC
by $\pm 2\sigma$.

The small peaking background in the $\mb$ signal region of $\fcp$ modes other
than $\jpsi\kstarz$ is neglected in the nominal analysis.
The effect of the fractions and their $CP$ asymmetries is
studied with MC simulation and is included in systematic errors.

Each physics parameter ($\taubz$, $\dmd$, $J/\psi K^{*0}$ polarization)
is also varied by its error; for $\dmd$, we also
use our result ($\dmd = \dmdcenter$ ps$^{-1}$), repeat the fit
and take the larger change as the systematic error.

Finally, we investigate the effects of interference between
CKM-favored and CKM-suppressed $B\to D$ transitions in
the $\ftag$ final state~\cite{Long:2003wq}.
A small correction to the PDF for the signal distribution
arises from the interference.
We estimate the amount of correction using the $\bzdslnu$
sample. We then generate MC pseudo-experiments
and make an ensemble test to obtain systematic biases
in $\sinbb$ and $|\lambda|$. We find that
the effect on $\sinbb$ is negligibly small, while
a possible shift in $|\lambda|$ becomes the largest
contribution to the systematic error.

Several checks on the measurement are performed.
Table~\ref{tab:check} lists the results obtained by applying the same analysis
to various subsamples.
\begin{table}
  \caption{Numbers of candidate events, $\Nev$,
    and values of $\sinbb$, $|\lambda|$
    for various subsamples (statistical errors only).}
  \begin{ruledtabular}
    \begin{tabular}{lrcc}
      Sample & \multicolumn{1}{c}{$\Nev$} & $\sinbb$ & $|\lambda|$ \\
      \hline
      $\jpsi\ks(\pip\pim)$    & 1997 & $0.67 \pm 0.08$ & $0.98 \pm 0.06$ \\
      $J/\psi \ks(\piz\piz)$  &  288 & $0.72 \pm 0.20$ & $1.18 \pm 0.27$\\
      $\psi(2S) \ks$          &  308 & $0.89 \pm 0.20$ & $0.94 \pm 0.14$\\
      $\chi_{c1}\ks$          &  101 & $1.54 \pm 0.49$ & $0.76 \pm 0.22$\\
      $\eta_c \ks$            &  217 & $1.32 \pm 0.28$ & $1.10 \pm 0.30$\\
      \cline{2-4}
      All with $\xi_f = -1$   & 2911 & $0.73 \pm 0.06$ & $0.99 \pm 0.05$ \\
      \hline
      $\jpsi\kl$              & 2332 & $0.77 \pm 0.13$ & $1.04 \pm 0.08$\\
      $\jpsi\kstarz(\ks\piz)$ &  174 & $0.10 \pm 0.45$ & $1.11 \pm 0.33$\\
      \hline
      $\ftag = \bz$ ($q=+1$)  & $\nevqp$ & $0.72 \pm 0.09$ & $0.89\pm0.09$ \\
      $\ftag = \bzb$ ($q=-1$) & $\nevqm$ & $0.74 \pm 0.08$ & $1.17\pm0.11$ \\
      \hline
      $0 < r \le 0.5$         & 2985 & $0.95 \pm 0.26$ & $1.18\pm0.22$ \\
      $0.5 < r \le 0.75$      & 1224 & $0.68 \pm 0.11$ & $1.11\pm0.09$ \\
      $0.75 < r \le 1$        & 1208 & $0.73 \pm 0.07$ & $0.95\pm0.05$ \\
      \hline
      Data set I (78 fb$^{-1}$) & 3013 & $0.72 \pm 0.07$ & $0.95\pm0.05$ \\
      Data set II (62 fb$^{-1}$)& 2404 & $0.74 \pm 0.09$ & $1.09\pm0.07$ \\
      \hline \hline
      All                     & 5417 & $\sinbbcenter\pm\sinbbstat$ 
                                          &$\lambdacenter\pm\lambdastat$\\
    \end{tabular}
  \end{ruledtabular}
\label{tab:check} 
\end{table}
All values are statistically consistent with each other.
\begin{figure}
  \includegraphics[width=0.6\textwidth,clip]{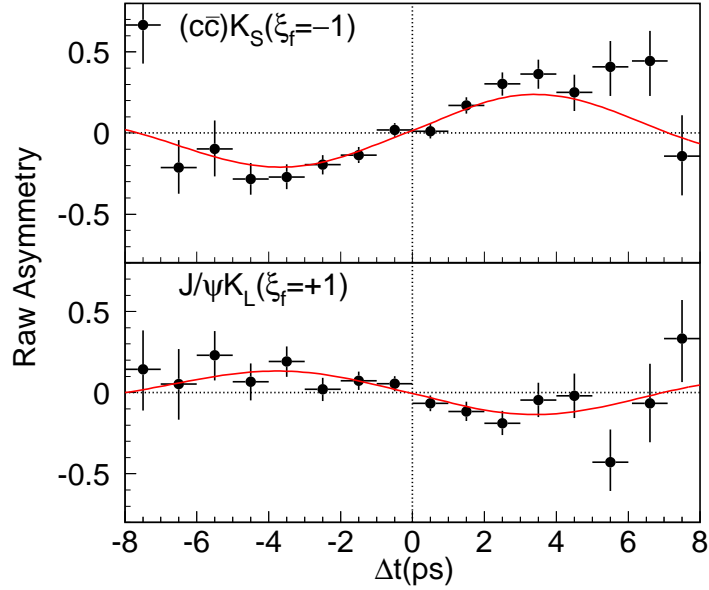}
  \caption{Raw asymmetries for $(c\overline{c})\ks$ ($\xi_f = -1$) (top) and
    $\jpsi\kl$ ($\xi_f = +1$) (bottom).
    The curves are the results of the global unbinned maximum-likelihood fit.}
\label{fig:rawasym}
\end{figure}
Figure~\ref{fig:rawasym} shows the raw asymmetries
and the fit results for $(c\overline{c})\ks$ (top) and $\jpsi\kl$ (bottom).
A fit to the non-$CP$ eigenstate modes
$\bz \to D^{*-}\ell^+\nu$ and $\jpsi\kstarz(\kp\pim)$,
where no asymmetry is expected, yields 
``$\sin2\phi_1$''$ = 0.012 \pm 0.013$(stat).

\section{Summary}
\label{sec:summary}

Using $152 \times 10^6$ $B\overline{B}$ pairs 
collected at the $\ufours$ resonance 
with the Belle detector at the KEKB asymmetric-energy $e^+e^-$ collider,
we have measured the $CP$-violation parameters
$\sinbb$ and $|\lambda|$, $B$ meson lifetimes and their ratio, and
the mixing parameter $\dmd$.
These are basic parameters of the standard model.
The results are summarized as follows:
\begin{eqnarray}
 \sinbb     &=& \sinbbresult,     \nonumber \\
 |\lambda|  &=& \lambdaresult,    \nonumber \\
 \taubz     &=& \taubzresult,     \nonumber \\
 \taubp     &=& \taubpresult,     \nonumber \\
 \taubratio &=& \taubratioresult, \nonumber \\
 \dmd       &=& \dmdresult.       \nonumber
\end{eqnarray}
All results are significant improvements in precision
from the previous measurements, and are in agreement
with the standard model expectations.
The significance of
the observed deviation from unity in the lifetime ratio 
exceeds five standard deviations for the first time by a single measurement.

%
\begin{acknowledgments}
We thank the KEKB group for the excellent operation of the
accelerator, the KEK Cryogenics group for the efficient
operation of the solenoid, and the KEK computer group and
the National Institute of Informatics for valuable computing
and Super-SINET network support. We acknowledge support from
the Ministry of Education, Culture, Sports, Science, and
Technology of Japan and the Japan Society for the Promotion
of Science; the Australian Research Council and the
Australian Department of Education, Science and Training;
the National Science Foundation of China under contract
No.~10175071; the Department of Science and Technology of
India; the BK21 program of the Ministry of Education of
Korea and the CHEP SRC program of the Korea Science and
Engineering Foundation; the Polish State Committee for
Scientific Research under contract No.~2P03B 01324; the
Ministry of Science and Technology of the Russian
Federation; the Ministry of Education, Science and Sport of
the Republic of Slovenia;  the Swiss National Science Foundation; the
National Science Council and
the Ministry of Education of Taiwan; and the U.S.\
Department of Energy.
\end{acknowledgments}



\begin{thebibliography}{99}
\bibitem{footnote:cc}
  In this paper the inclusion of charge conjugate state is implied.
\end{thebibliography}
\end{document}